\documentclass[
 prx, reprint,
superscriptaddress,
 amsmath,amssymb,
 aps
]{revtex4-2}
\usepackage{color}
\usepackage{booktabs}
\usepackage{makecell}
\usepackage{tabularx}
\usepackage{array}
\usepackage{graphicx}
\usepackage{subfigure}
\usepackage{stfloats}
\usepackage{caption}
\usepackage{multirow}
\usepackage{mathtools}
\usepackage{dcolumn}
\usepackage{bm}
\usepackage{algorithm}
\usepackage{algorithmic}
\usepackage[colorlinks,
            linkcolor=blue,
            anchorcolor=blue,
            citecolor=blue,
            ]{hyperref}
\usepackage[T1]{fontenc}

\begin{document}

\title{Detecting Entanglement by Pure Bosonic Extension}

\author{Xuanran Zhu}
\thanks{These authors contributed equally to this work}
\affiliation{Department of Physics, The Hong Kong University of Science and Technology, Clear Water Bay, Kowloon, Hong Kong, China}

\author{Chao Zhang}
\thanks{These authors contributed equally to this work}
\affiliation{Department of Physics, The Hong Kong University of Science and Technology, Clear Water Bay, Kowloon, Hong Kong, China}

\author{Chenfeng Cao}
\affiliation{Department of Physics, The Hong Kong University of Science and Technology, Clear Water Bay, Kowloon, Hong Kong, China}

\author{Youning Li}
\affiliation{College of Science, China Agricultural University, Beijing, 100080, People's Republic of China}

\author{Yiu Tung Poon}
\affiliation{Department of Mathematics, Iowa State University, Ames, Iowa 50011, USA}

\author{Bei Zeng}
\email{zengb@ust.hk}
\affiliation{Department of Physics, The Hong Kong University of Science and Technology, Clear Water Bay, Kowloon, Hong Kong, China}

\date{\today}
\begin{abstract}
In the realm of quantum information theory, the detection and quantification of quantum entanglement stand as paramount tasks. The relative entropy of entanglement (REE) serves as a prominent measure of entanglement, with extensive applications spanning numerous related fields. The positive partial transpose (PPT) criterion, while providing an efficient method for the computation of REE, unfortunately, falls short when dealing with bound entanglement. In this study, we propose a method termed "pure bosonic extension" to enhance the practicability of $k$-bosonic extensions, which approximates the set of separable states from the "outside", through a hierarchical structure. It enables efficient characterization of the set of $k$-bosonic extendible states, facilitating the derivation of accurate lower bounds for REE. Compared to the Semi-Definite Programming (SDP) approach, such as the symmetric/bosonic extension function in QETLAB, our algorithm supports much larger dimensions and higher values of extension $k$.
\end{abstract}

\maketitle

\section{Introduction}
\label{intro}
Entanglement, a feature of quantum mechanics that was first described by Einstein, Podolsky, and Rosen \cite{PhysRev.47.777}, stands as one of the most fascinating aspects of the field. As quantum information theory has emerged, entanglement has come to be recognized not just as a phenomenon, but as a resource in many quantum information tasks, ranging from quantum cryptography \cite{PhysRevLett.67.661} and quantum teleportation \cite{PhysRevLett.70.1895}, to quantum computation \cite{PhysRevLett.86.5188}. Despite ongoing efforts to establish a universal criterion for detecting entanglement, it continues to be an unresolved challenge, confirmed to be NP-hard \cite{10.1145/780542.780545}.

For a given bipartite system $AB$ with $\text{dim}(\mathcal{H}_A)=d_A$ and $\text{dim}(\mathcal{H}_B)=d_B$, a state $\rho_{AB}$ is separable if it can be written in a convex combination form
\begin{equation}
	\rho_{AB}=\sum_{i}p_i \rho_A^{(i)}\otimes \rho_B^{(i)},
\end{equation}
where $\sum_i p_i=1,p_i\geq 0$, $\rho_A^{(i)}$ and $\rho_B^{(i)}$ are local density matrices in Hilbert spaces $\mathcal{H}_A$ and $\mathcal{H}_B$, respectively. Otherwise, it is entangled \cite{Werner_1989}.

Among many methods for entanglement detection and quantification, relative entropy of entanglement (REE) $E_R(\rho)$ is one important quantity \cite{vedral2002role,miranowicz2004comparative}, which is defined as
\begin{equation}
	E_R(\rho)=\min_{\rho^{\prime} \in \mathrm{SEP}} S(\rho \vert \vert \rho^{\prime})=\min_{\rho^{\prime} \in \mathrm{SEP}} \text{Tr}[\rho \text{ln}\rho -\rho\text{ln}\rho^{\prime}],
\end{equation}
where $\mathrm{SEP}$ denotes the set of separable states. $E_R(\rho)$ can be considered as the optimal distinguishability of the state $\rho$ from separable states. The calculation of $E_R(\rho)$ necessitates the resolution of an optimization problem \cite{Fawzi_2018, Girard_2014}, a task rendered challenging owing to the imperative of characterizing the separable set.

In practice, the computation of $E_R(\rho)$ can be achieved by determining its upper and lower bounds. Various methods exist for obtaining these bounds. For example, one can approximate the separable set from the "inside" using convex hull approximation (CHA) \cite{lu2018separability,hou2020upper}, which results in an upper bound $E^u_R(\rho)$. On the other hand, the Semi-Definite Programming (SDP) method \cite{Vandenberghe_1996} can be employed to determine a lower bound $E^l_R(\rho)$ by optimizing over the set of states with positive partial transpose (PPT) \cite{PhysRevLett.77.1413, HORODECKI19961}, which approximates the separable set from the "outside". However, it is well-recognized that a gap exists between the boundaries of the PPT set and the separable set, where bound entanglement resides \cite{horodecki1998mixed}. Within this region, the PPT criterion does not yield accurate results.

Alternatively, $k$-symmetric/bosonic extension also approximates the separable set from the "outside" \cite{PhysRevLett.88.187904}. Here, we mainly focus on the set of $k$-bosonic extendible states $\bar{\Theta}_k$ which is a convex subset of the set of $k$-symmetric extendible states $\Theta_k$. A bipartite state $\rho_{AB}$ is considered $k$-bosonic extendible if there exists a global quantum state $\rho_{AB_1 B_2 ...B_k}$, which is supported on the symmetric subspace of $B_1 B_2 ...B_k$, and preserves the marginals on $AB_i$ equal to $\rho_{AB}$. A hierarchy structure has been proved in the set $\bar{\Theta}_k$, i.e., $\bar{\Theta}_{k+1}\subset \bar{\Theta}_{k}$, and $\bar{\Theta}_\infty=\mathrm{SEP}$ \cite{Doherty_2004}, explaining how this method approaches SEP asymptotically. Owing to their convex nature, determining whether a quantum state belongs to $\bar{\Theta}_k$ can also be formulated as an SDP problem \cite{PhysRevA.80.052306}, which can then be resolved by tools like QETLAB \cite{qetlab}. In principle, $k$-symmetric/bosonic extension can yield a superior lower bound for $E_R(\rho)$ compared to the one obtained via the PPT criterion, particularly when $k$ reaches a sufficient number.

It is worth noting that the $k$-symmetric/bosonic extension bears relevance to the $N$-representability problem \cite{Chen_2014}, which is QMA-complete \cite{PhysRevLett.98.110503}. This implication suggests that, even when assisted by large-scale fault-tolerant quantum computers, efficiently solving this problem in the worst-case scenario remains improbable. As a result, the complexity of solving SDP problems for the $k$-symmetric/bosonic extension escalates rapidly as the value of $k$ increases. The practicability of tools like QETLAB is significantly limited, as they can only address problems of small dimensions and few extensions.

\begin{figure}[H]
	\centering
	\includegraphics[width=1\linewidth]{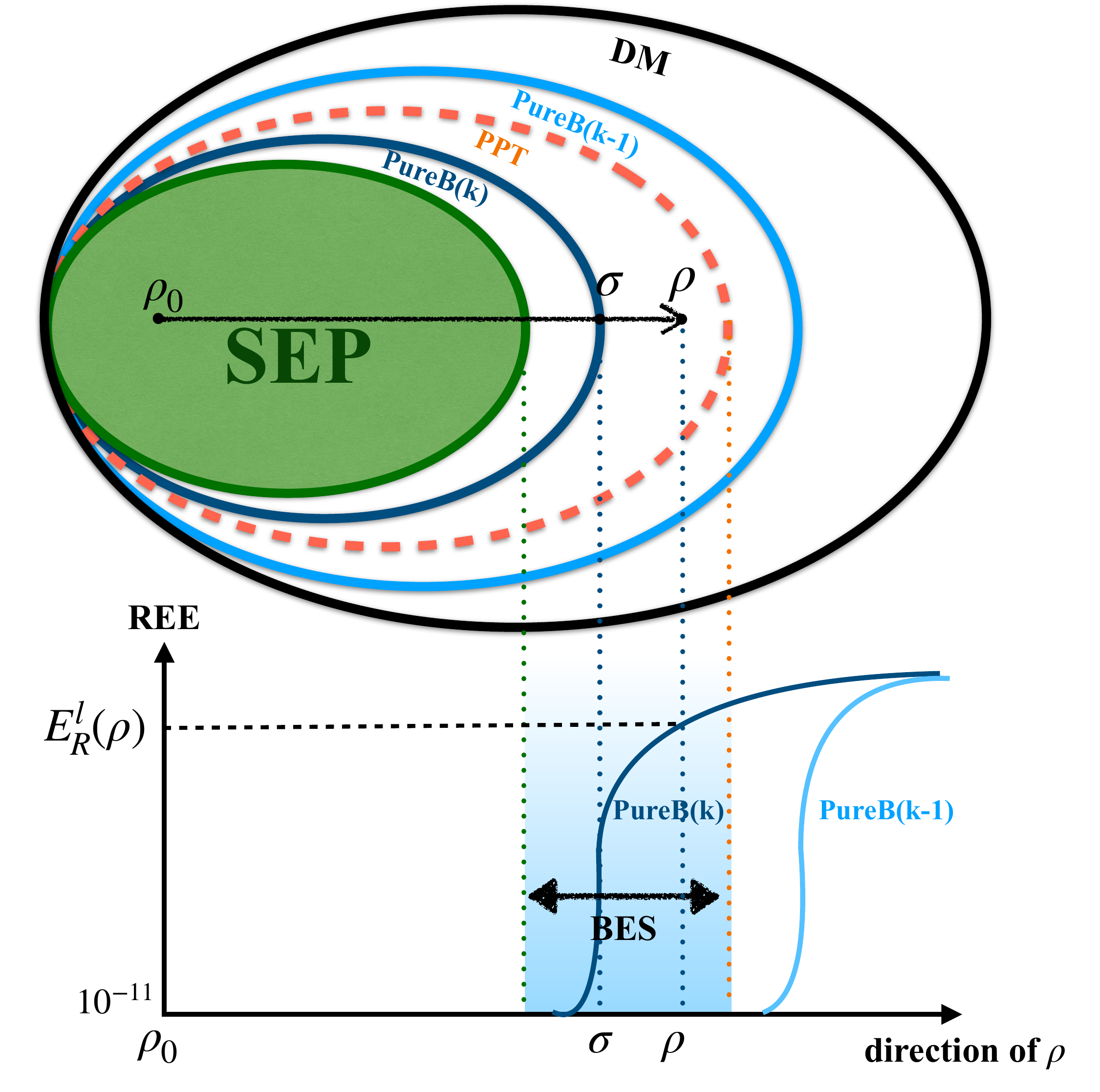}
	\caption{The figure illustrates the behavior of our PureB-ext method in density matrix space, which approximates the separable set (green ellipse) from outside. The lines sketch the boundaries of the separable set (SEP), $k$-bosonic extendible set with pure pre-images (PureB($k$)), the set of states with positive partial transpose (PPT), and density matrices (DM). From the maximally mixed state $\rho_0$, going along the direction of a quantum state $\rho$ in density matrix space, we can see a clear and sharp transition in REE across the boundary state $\sigma$, which can then give a nice lower bound $E^{l}_R(\rho)$ even if $\rho$ is a bound entangled state.}
	\label{algorithm_demo}
\end{figure}

In this work, we propose a new method to characterize the $k$-bosonic extendible set $\bar{\Theta}_k$, focusing on those $k$-bosonic extendible states with \textit{pure} pre-images. Through our investigation of the numerical range \cite{li2000convexity} of $\bar{\Theta}_k$, we find that for generic states on the \textit{boundary}, their pre-images are always pure when $k$ exceeds a certain transition value $k^{\star}$ related to the dimensions of the subsystems $A$ and $B$. Surprisingly, we further observe that $k$-bosonic extendible states with pure pre-images can characterize $\bar{\Theta}_k$ quite effectively even for the \textit{interior} states, giving a clear and sharp transition across the boundary. This is what we refer to as the "pure bosonic extension", denoted by PureB-ext, as depicted in Fig. \ref{algorithm_demo}.

By parameterizing pure $k$-bosonic states $\vert \psi_{AB_1...B_k}\rangle$ and employing the gradient backpropagation technique, we can obtain satisfactory lower bounds for REE in various cases, including the region of bound entanglement. Compared to the traditional SDP approach, such as the symmetric/bosonic extension function in QETLAB, our algorithm can handle much larger dimensions and provide more extensions.

This paper is organized as follows: Sec.~\ref{Sec:preliminaries} lays the groundwork by introducing necessary definitions and facts, while elucidating the motivation behind our research. The pure bosonic extension (PureB-ext) methodology is introduced in Sec.~\ref{Sec:method}, wherein we delineate the computation process for the lower bound of REE through this method and validate the effectiveness of PureB-ext with numerical evidence. Sec.~\ref{Sec:results} presents our calculation results in diverse scenarios, compared with results obtained using other established methods. Finally, Sec.~\ref{Sec:discussions} hosts further discussions and provides forward-looking perspectives on the topic.

\section{Preliminaries}\label{Sec:preliminaries}

\subsection{Vectorization of density matrix space}

To represent a $d$-by-$d$ density matrix $\rho\in\mathbb{C}^{d\times d}$ as a real vector $\vec{\rho}\in\mathbb{R}^{d^2-1}$, we can find a Hermitian orthogonal basis that contains identity such that $\rho$ can be expanded in such a basis with real coefficients. These coefficients can be viewed as the real space coordinates of the quantum state on the given basis. For example, the Pauli matrices are commonly used for $2$-by-$2$ density matrices. For higher dimensions, we can use generalized Gell-Mann matrices $\{\lambda_i\}$ \cite{Bertlmann_2008} , satisfying
\[
\text{Tr}[\lambda_i]=0, \quad \text{Tr}[\lambda_i\lambda_j]=2\delta_{ij}, \quad \lambda_i^\dagger=\lambda_i.
\]
Then any $d$-by-$d$ density matrix $\rho$ can be written as
\begin{equation}
	\rho=\rho_0 + \vec{\rho} \cdot \vec{\lambda}, \label{eq:dm-vector-representation}
\end{equation}
where $\rho_0=\mathbb{I}/d$ is the maximally mixed state and $\vec{\rho}=(x_1,x_2,\dots,x_{d^2-1}) \in \mathbb{R}^{d^{2}-1}$ satisfies
\[
x_i =\frac{1}{2} \text{Tr}[\rho\lambda_i], \quad i=1,2,\dots,d^2-1.
\]

With the above decomposition, every density matrix can be mapped to a vector in a density matrix space as shown in Fig.~\ref{algorithm_demo}, where the maximally mixed state $\rho_0$ corresponds to the origin point $\vec{0}$. The distance between two density matrices $\rho_1$ and $\rho_2$ can be defined as the Euclidean distance between their vectorized forms:
\[
D(\rho_1,\rho_2) = \Vert \vec{\rho_1}-\vec{\rho_2}\Vert_2,
\]
where $\rho_1 = \rho_2$ if and only if $D(\rho_1,\rho_2)=0$.
The length of a quantum state $\rho$ can then be defined as $\Vert\vec{\rho}\Vert_2$, i.e., the Euclidean distance from the origin point, which is related to the purity of the state $\rho$:
\[
\gamma(\rho) = \text{Tr}[\rho^2] = \frac{1}{d} + 2\Vert\vec{\rho}\Vert_2^2.
\]
Since the purity cannot exceed $1$, we can further obtain the outermost boundary of density matrix space where $\Vert\vec{\rho}\Vert_2 \leq \sqrt{\frac{(d-1)}{2d}}$.

Given a set of quantum states $\mathcal{S}$, we can also define the distance between a state $\rho$ and $\mathcal{S}$ as
\begin{equation}
	\label{eq:distance-to-set}
	D(\rho,\mathcal{S}) = \min_{\rho'\in \mathcal{S}} D(\rho,\rho').
\end{equation}
If $\rho \in \mathcal{S}$, then $D(\rho,\mathcal{S})=0$. Otherwise, $D(\rho,\mathcal{S})>0$.

\subsection{The set of $k$-extendible states}

We recall some facts about $k$-(symmetric/bosonic) extendible states and their relationship to separability.

A bipartite state $\rho_{AB}$ is said to be $k$-symmetric extendible if there exists a global state $\rho_{AB_1 B_2...B_k}$ such that for any $i,j \in \{1,2,...,k\}$
\begin{equation}
	\begin{aligned}
		\operatorname{Tr}_{B_2\cdots B_k}\left[\rho_{A B_1 B_2 \cdots B_k}\right] & =\rho_{A B}, \\
		\left(\mathbb{I}_A \otimes P_{i j}\right) \rho_{A B_1 B_2 \cdots B_k} \left(\mathbb{I}_A \otimes P_{i j}\right)& =\rho_{A B_1 B_2 \cdots B_k}.
		\end{aligned}
\end{equation}
Here, $P_{ij}$ is a permutation operator that exchanges the $i$-th and $j$-th subsystems in $B_1 B_2\dots B_k$, $\text{dim}(\mathcal{H}_A)=d_A$ and $\text{dim}(\mathcal{H}_{B_i})=d_B$. The set of all $k$-symmetirc extendible states, denoted by $\Theta_k$, is convex with a hierarchy structure $\Theta_{k+1}\subset \Theta_{k}$. Moreover, when $k \rightarrow \infty$, $\Theta_k$ converges exactly to the set of separable states \cite{PhysRevLett.88.187904}.

The $k$-bosonic extendible set $\bar{\Theta}_k$ is a convex subset of $\Theta_k$, which further requires the global state $\rho_{AB_1 B_2...B_k}$ is supported on the symmetric subspace of $B_1 B_2\dots B_k$. Similarly, a hierarchy structure arises in the set $\bar{\Theta}_k$, i.e., $\bar{\Theta}_{k+1}\subset \bar{\Theta}_{k}$, and $\bar{\Theta}_{\infty}=\text{SEP}$. In the following, we denote the set of all states $\rho_{A B_1 B_2 \cdots B_k}$ supported on the symmetric subspace as $\mathcal{B}_k$. Let $\varphi_k = \operatorname{Tr}_{B_2...B_k}$, we know that $\bar{\Theta}_k=\varphi_k(\mathcal{B}_k)$.

The set of $k$-extendible states is known to be closely related to the ground state of some $(k+1)$-body Hamiltonians \cite{Zeng_2019}. To elucidate this relationship, we focus on the bosonic extension and adopt the generalized Gell-Mann matrices $\{ \lambda_i \}$ as our basis.  For any $\rho \in \bar{\Theta}_k$, there exists a $\rho^{\prime} \in \mathcal{B}_k$ such that
\[
	\operatorname{Tr}[\lambda_i\rho]=\operatorname{Tr}[\lambda_i\varphi_k(\rho^{\prime})]=\operatorname{Tr}[\varphi_k^{*}(\lambda_i)\rho^{\prime}],
\]
where $\varphi_k^{*}$ is the adjoint map of $\varphi_k$. The expression $\operatorname{Tr}[\lambda_i\rho]$ is connected with an important concept known as the numerical range (NR) as follows:
\begin{equation}
	\begin{aligned}
		W(\{\lambda_i\})&=\{(\operatorname{Tr}[\lambda_1\rho],...,\operatorname{Tr}[\lambda_{d^2-1}\rho]):\rho \in \bar{\Theta}_k\}\\
		&=\{(\operatorname{Tr}[A_1\rho^{\prime}],...,\operatorname{Tr}[A_{d^2-1}\rho^{\prime}]):\rho^{\prime} \in \mathcal{B}_k\},
	\end{aligned}
	\label{eq:numerical-range}
\end{equation}
where $A_i=\varphi_k^{*}(\lambda_i)$ and $d=d_Ad_B$. $W$ not only represents the joint algebraic numerical range of $\bar{\Theta}_k$ on the basis $\{\lambda_i\}$ but also can be considered as the projection of $\mathcal{B}_k$ onto the low-dimensional subspace constructed by $\{A_i\}$.

In practice, the $k$-bosonic extendible set $\bar{\Theta}_k$ can be explored through its NR, due to the one-to-one correspondence depicted in Eq.~(\ref{eq:dm-vector-representation}). Here, the NR can be viewed as a graphical representation of the given set, with the boundary of the set being of prime importance. Based on the studies of numerical ranges \cite{li2000convexity}, it is known that the extreme point on the boundary of $W$ can be determined by the ground state of some Hamiltonian $H=-\sum_i n_iA_i$, where $\hat{n}=(n_1,n_2,...,n_{d^2-1})\in\mathbb{R}_{d^2-1}$ represents the unit normal vector on the boundary. Those ground states belong to $\mathcal{B}_k$ since the bosonic symmetry of the given Hamiltonian $H$. However, if for some direction $\vec{n^{\star}}$, Hamiltonian $H^{\star}$ exhibits degeneracy in its ground states, that implies the presence of a flat boundary which may not be obtained by pure states in $\mathcal{B}_k$. 

\subsection{Degeneracy contraction on the boundary}
\label{sec:degeneracy-contraction}
According to the Ref.~\cite{friedland1976subspaces}, there is a best possible bound for the existence of degeneracy in a matrix subspace, which is given by the following theorem.

\textbf{Theorem 1}. Let $\mathcal{H}$ be a $m$-dimensional subspace in the space of $n\times n$ Hermitian matrices. If

\[ m \geq (r-1)(2n-r+1), \]

then $\mathcal{H}$ contains a nonzero matrix $H^{\star}$ such that the greatest eigenvalues of $H^{\star}$ is at least of multiplicity $r$, where $2 \leq r \leq n-1$.

In essence, when $(d_A d_B)^2-1 \geq 2d_Ad_k-1$, where $d_k$ is the dimension of the symmetric subspace in $B_1\dots B_k$, degeneracy can invariably be detected. However, as the dimension of the symmetric subspace increases, the degeneracy tends to disappear generically. The transition dimension for the symmetric subspace can be \textit{estimated} around $\frac{1}{2}d_A d_B^2$ (the exact value depends on the specific structure of the given matrix subspace, in our case, they have bosonic symmetry).

Consequently, when $k>k^{\star}$, for randomly selected states on the boundary of $\bar{\Theta}_k$, they always have pure pre-images $|\psi_{AB_1...B_k}\rangle \in \mathcal{B}_k$. But it is worth noting, we have not excluded the possibility of degeneracy in higher dimensions. What we emphasize here is that in generic cases, there are always no degeneracies.

\section{Methodology}\label{Sec:method}

As informed by the discussion in the preceding section, we recognize that the marginals of pure $k$-bosonic states $|\psi_{AB_1...B_k}\rangle \in \mathcal{B}_k$ on $AB_i$ can \textit{almost} characterize the boundary of $\bar{\Theta}_k$ when $k$ surpasses a particular transition value $k^{\star}$. 

We then proceed to investigate the above phenomenon through some numerical analysis. For simplicity, We designate the approach which only considers the states with pure pre-images as the "pure bosonic extension", denoted as PureB-ext.

\subsection{Parameterization}
The generalized Dicke states form a complete orthogonal basis for the symmetric subspace of $B_1\dots B_k$, which can be written as
\[ \vert D^{k}_{\vec{w}}\rangle=\left(\frac{k!}{w_0!w_1!...w_{d_B-1}!}\right)^{-1/2}\sum_{\text{wt}(x)=\vec{w}}\vert x\rangle, \]
where $\text{wt}(x)=\vec{w}$ means the number of particles on $i$-th energy level is $w_i(i=0,1,...,d_B-1)$, satisfying the conservation of particle number $\sum_i w_i = k$. For example,
\[ \vert D^4_{1,3}\rangle=\frac{1}{2}(\vert 0111\rangle+\vert 1011\rangle+\vert 1101\rangle+\vert 1110\rangle).\]
And the dimension of the symmetric subspace is
\[ d_k :=\mathrm{dim}(\mathcal{H}_{B_1...B_k})=\binom{k+d_B-1}{d_B-1}=\frac{(k+d_B-1)!}{(d_B-1)!k!}, \]
e.g., for the qubit case, the dimension is $k+1$. Furthermore, a recurrence relation can be obtained for generalized Dicke states:
\begin{equation}
	\label{recurrence}
	\vert D^{k}_{w_0,...,w_{d_B-1}}\rangle=\sum_{i=0}^{d_B-1}\sqrt{\frac{w_i}{k}}\vert i\rangle\otimes\vert D^{k-1}_{w_0,...,w_i-1,...,w_{d_B-1}}\rangle.
\end{equation}

Based on the above knowledge, we can parameterize the pure $k$-bosonic state $\vert \psi_{AB_1...B_k}\rangle$ as the following form
\begin{equation}
	\label{eq:psi-ABn}
	\vert \psi_{AB_1...B_k}\rangle=\sum_{i,\vec{w}} c_{i,\vec{w}}\vert i\rangle\otimes\vert D^{k}_{\vec{w}}\rangle.
\end{equation}
Then the corresponding marginal state on $AB_i$ can be obtained as
\begin{equation}
	\label{eq:RDM-dicke}
	\begin{aligned}
		\rho_{AB}=&\text{Tr}_{B_2...B_k}[\vert \psi_{AB_1...B_k}\rangle\langle \psi_{AB_1...B_k}\vert]\\
		=&\sum_{i,j,\vec{w},\vec{w}^{\prime},r,s}c_{i,\vec{w}}c_{j,\vec{w}^{\prime}}^{*}B^k_{\vec{w}\vec{w}^{\prime}rs}\vert i\rangle\langle j\vert\otimes \vert r\rangle \langle s \vert,
	\end{aligned}
\end{equation}
where the marginal bosonic coefficient $B^k_{\vec{w}\vec{w}^{\prime}rs}$ can be derived from the recurrence relation in Eq.~(\ref{recurrence}):
\begin{align*}
	B_{\vec{w}\vec{w}^{\prime}rs}^{k}&=\left\langle r\right|\mathrm{Tr}_{B_{2}\cdots B_{k}}\left[\left|D_{\vec{w}}^{k}\right\rangle \left\langle D_{\vec{w}'}^{k}\right|\right]\left|s\right\rangle \\&=\frac{1}{k}\sqrt{w_{r}w_{s}^{\prime}}\delta_{w_{0},w_{0}^{\prime}}...\delta_{w_{r}-1,w_{r}^{\prime}}...\delta_{w_{s},w_{s}^{\prime}-1}...\delta_{w_{d-1},w_{d-1}^{\prime}}.
\end{align*}

\subsection{Analysis}
With the parameterization of $\vert \psi_{AB_1...B_k}\rangle$, we can efficiently characterize pure states in $\mathcal{B}_k$ and subsequently investigate the behavior of the PureB-ext across the boundary of $\bar{\Theta}_k$. 

We randomly select directions $\hat{\rho}$ in the density matrix space and evaluate the Euclidean distance defined in Eq.~(\ref{eq:distance-to-set}) between the given state $\rho$ and the marginals of all pure states $\vert \psi_{AB_1...B_k}\rangle$. This evaluation is conducted repeatedly along the selected directions, starting from the maximally mixed state $\rho_0=I/d$ and ending at the quantum state boundary. The boundary of $k$-bosonic extendible set can be calculated exactly by SDP in QETLAB \cite{qetlab} with the parameter \textit{bos=1}.

\begin{figure}[H]
	\centering
	\includegraphics[width=\linewidth]{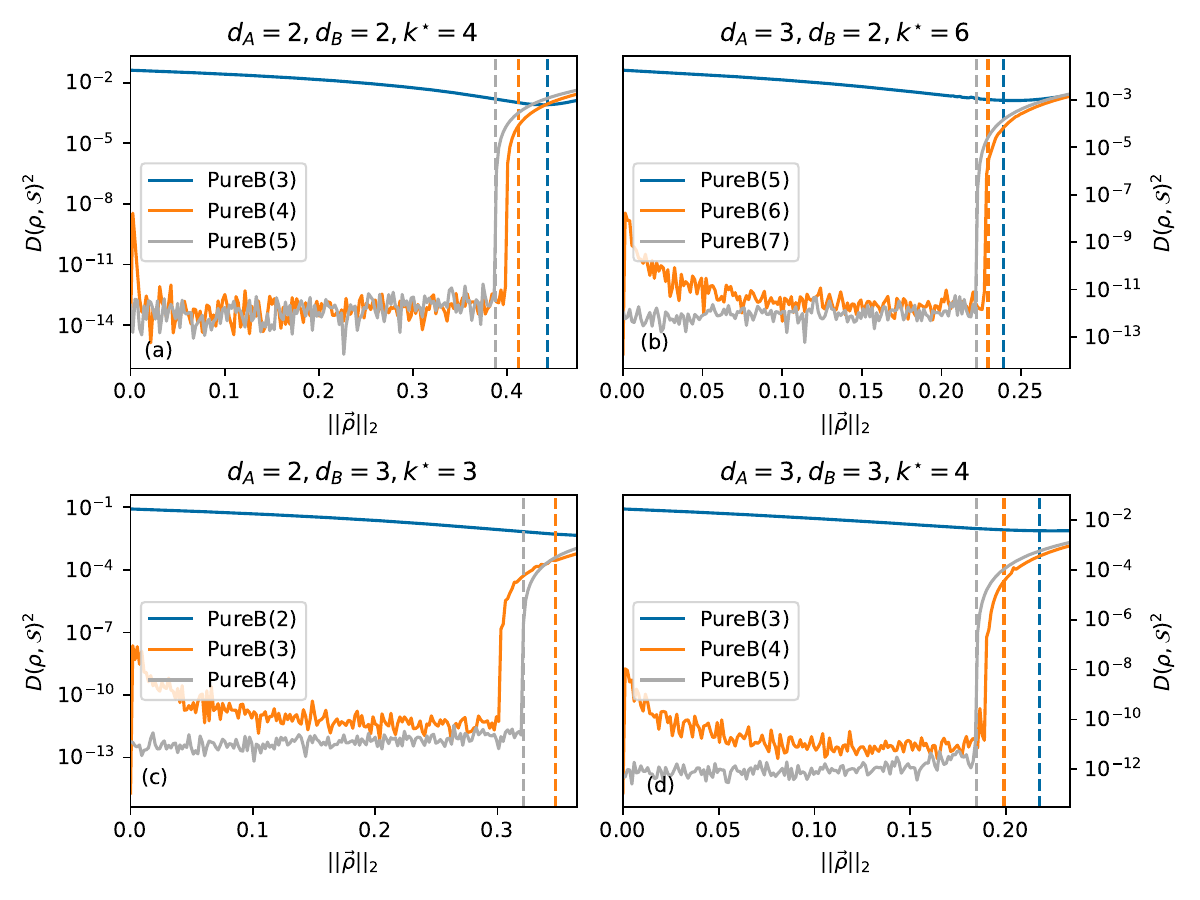}
	\caption{Squared Euclidean distance $D(\rho, \mathcal{S})^2$ for various extension numbers. Here, $\mathcal{S}$ denotes the set of marginals of pure $k$-bosonic states. The solid lines are the values computed by PureB-ext and the dashed vertical lines correspond to the exact $k$-bosonic boundaries from SDP. The sharp transition implies the boundary of certain $\mathcal{S}$. Transition values $k^{\star}$ can be computed by Eq.~(\ref{eq:transition-value}), which is consistent with the numerical results as shown in the subtitles.}
	\label{fig:critical-point-kstar}
\end{figure}

The numerical results are shown in Fig.~\ref{fig:critical-point-kstar} with bipartite system $2\otimes 2,3\otimes 2,2\otimes 3,3\otimes 3$ in the subfigure (a/b/c/d) correspondingly. When $k$ is relatively small, e.g. $k=3$ in subfigure (a) for $2 \otimes 2$ case, PureB-ext cannot provide correct information about the exact boundary. When $k$ is large enough, e.g. $k=5$, the sharp change of Euclidean distance given by the PureB-ext matches with the exact $k$-bosonic boundary quite well. Therefore, $k^{\star}=4$ can be considered as a transition value for $2\otimes 2$ system. This phenomenon is consistently observed in all four subfigures. Additionally, we note that the marginals of $\vert \psi_{AB_1...B_k}\rangle$ not only establish the boundaries of $\bar{\Theta}_k$ but also effectively characterize the interior states when $k$ exceeds the transition value. A simple explanation is that when $k$ is large enough, the projection of those pure states in $\mathcal{B}_k$ onto $\bar{\Theta}_k$ is sufficiently dense. One typical interior state is the maximally mixed state $\rho_0$ in $2\otimes 2$, which has the pure pre-image as
\[
	\begin{aligned}
		\left|\psi_{A B_1\cdots B_4}\right\rangle= & |0\rangle \otimes\left(b\left|D_0^4\right\rangle-a\left|D_2^4\right\rangle+b\left|D_4^4\right\rangle\right) \\
		& +c|1\rangle \otimes\left(\left|D_1^4\right\rangle+\left|D_3^4\right\rangle\right),
		\end{aligned}
\]
where $a=\frac{1}{2\sqrt{2}}, b= \frac{\sqrt{3}}{4}, c= \frac{1}{2}$. For other dimensions, the analytical form of the pure pre-image of $\rho_0$ can be written in the same way. However, the mechanism of the PureB-ext for interior states is still unclear and needs further investigation.

In conclusion, we conjecture that the marginals of pure $k$-bosonic states can effectively characterize the set of $k$-bosonic extendible states $\bar{\Theta}_k$ (for both boundary and interior states) when $k>k^{\star}=\lfloor k^{\prime}+1 \rfloor$, where $k^{\prime}$ satisfies
\begin{equation}
	\binom{k^{\prime}+d_B-1}{d_B-1}=\frac{1}{2}d_A d_B^2.
\label{eq:transition-value}
\end{equation}

\subsection{Optimization}
With the pure bosonic extension, we can effectively characterize the $k$-bosonic extendible set $\bar{\Theta}_k$ and efficiently compute the objective functions over $\bar{\Theta}_k$. One famous application is that we can calculate the relative entropy of a quantum state $\rho$ over $\bar{\Theta}_k$ to obtain a lower bound for the relative entropy of entanglement (REE):
\begin{equation}
	\label{eq:REE}
	E^l_R(\rho)=\min_{\rho_{AB}\in \bar{\Theta}_k}S(\rho||\rho_{AB}),
\end{equation}
which can be calculated efficiently by optimizing the coefficients $\{ c_{i,\vec{w}} \}$ of pure $k$-bosonic state $\vert \psi_{AB_1...B_k}\rangle$ in Eq.~(\ref{eq:psi-ABn}) whose size is $O(d_A k^{d_B-1})$, much smaller than the dimension of $\mathcal{B}_k$. As the value of $k$ increases, the accuracy of the lower bounds improves. A variational algorithm can be summarized below.

\begin{algorithm}[H]
	\renewcommand{\algorithmicrequire}{\textbf{Input:}}
	\renewcommand{\algorithmicensure}{\textbf{Output:}}
	\caption{REE based on pure bosonic extension}
	\label{alg}
	\begin{algorithmic}[1]
	\REQUIRE $\rho \in \mathcal{H}_A \otimes \mathcal{H}_B$, tolerance $\epsilon$, extension number $k$
	\STATE Initialization: generate $\{c_{i,\vec{w}} \}$ randomly
	\STATE \textbf{repeat}
	\STATE $\rho_{AB}=\sum_{i,j,\vec{w},\vec{w}^{\prime},r,s}c_{i,\vec{w}}c_{j,\vec{w}^{\prime}}^{*}B^k_{\vec{w}\vec{w}^{\prime}rs}\vert i\rangle\langle j\vert\otimes \vert r\rangle \langle s \vert$
	\STATE $E_R^{l}=S(\rho \vert\vert \rho_{AB}),\nabla E_R^{l}(\text{gradient backpropagation)}$
	\STATE Update $\{c_{i,\vec{w}}\}$ by gradient descent
	\STATE \textbf{until} $ E_R^{l}$ converges up to tolerance $\epsilon$
	\ENSURE $ E_R^{l},\{c_{i,\vec{w}}\}$
	\end{algorithmic}
\end{algorithm}

The optimization can be divided into the forward pass and the backward pass. The forward pass starts with the normalization of the trainable variables
\[ c_{i,\vec{w}}=\frac{\tilde{c}_{i,\vec{w}}}{\sum_{i,\vec{w}}{\left| \tilde{c}_{i,\vec{w}} \right|^2}}, \]
where $c_{i,\vec{w}}$ on the left-hand side is the amplitude coefficient to construct the pure $k$-bosonic state in Eq.~(\ref{eq:psi-ABn}) and $\tilde{c}_{i,\vec{w}}$ are unnormalized and free-tuned parameters. Then, the reduced density matrix (RDM) is evaluated using Eq.~(\ref{eq:RDM-dicke}) instead of explicitly constructing the generalized Dicke basis. To perform matrix logarithm for computing REE, we adopt the Padé approximation \cite{fawzi2019semidefinite} with hyper-parameters $m=8$ and $k=6$. 
\begin{align*}
	\log \left( X \right) &\approx 2^kr_m\left( X^{1/2^k} \right)\\
	r_m\left( X \right) &\coloneqq \sum_{j=1}^m{w_j\frac{X-1}{t_j\left( X-1 \right) +1}}
\end{align*}
where $w_j$ and $t_j$ are the weights and the nodes of the Gauss-Legendre quadrature. With these techniques, the relative entropy of some given density matrix $\rho$ with respect to the RDM of the pure bosonic state $\vert \psi_{AB_1...B_k}\rangle$ can be calculated efficiently.

In the backward pass, the gradient of the relative entropy with respect to the trainable parameters $\tilde{p}_{i,\vec{w}}$ can be calculated automatically by the PyTorch framework \cite{pytorch}. Most operations above can be back-propagated in the PyTorch framework except the matrix square root used in the Padé approximation. The chain rule for the matrix square root requires solving the Sylverster equation \cite{al2013computing} for which we use the algorithm provided in the SciPy package \cite{2020SciPy-NMeth}

With all gradients of the trainable parameters obtained after the backward pass, the limited-memory BFGS algorithm implemented in the SciPy package is applied with the convergence tolerance $10^{-10}$. For such a non-linear optimization which might contain many local minimums, we run the optimization program several times with different initial values. Most numerical results presented in this work are re-run three times with the hope that the global minimum could be found.

\section{Results}\label{Sec:results}
In this section, we mainly focus on the feasibility of our algorithm for detecting entanglement in different situations and compare it with other well-known algorithms, including the PPT criterion \cite{PhysRevLett.77.1413, HORODECKI19961}, CHA method \cite{lu2018separability,hou2020upper}, and the symmetric/bosonic extension function in QETLAB \cite{qetlab}. Relative entropy of entanglement (REE) is used as the entanglement measure.

\subsection{States with analytically known REE}

We start with the famous Werner states for $d \otimes d$ bipartite systems. They have the form
\[ \rho_{W}(\alpha)=\frac{1}{d^2-d\alpha}(I_{d^2}-\alpha F), \]
where $F = \sum_{ij}\vert ij\rangle \langle ji\vert$ is a swap operator. Whether a Werner state is entangled can be given by the PPT criterion. In other words, when we calculate $E_R$ for Werner states, optimization over either a separable set or a PPT set will give the same value. $E_R$ also has analytical form \cite{vollbrecht2001entanglement} as the following
\[ E_R(\rho_W(\alpha))=
\begin{cases}
  0, & \text{if}\ \alpha <\frac{1}{d} \\
  S(\rho_W(\alpha)\vert\vert\rho_W(\frac{1}{d})), & \text{otherwise}
\end{cases}. \]

We then consider $d\otimes d$ isotropic states, which have the form
\[ \rho_{I}(\alpha)=\frac{1-\alpha}{d^{2}} I_{d^{2}}+\alpha\left|\psi_{+}\right\rangle\left\langle\psi_{+}\right|, \]
where $\left|\psi_{+}\right\rangle=\frac{1}{\sqrt{d}} \sum_{j}|j j\rangle$, i.e., maximally entangled state. Similar to the case of Werner states, PPT criterion or analytical method \cite{PhysRevA.60.179} can be used to calculate $E_R$. The analytical form follows as
\[ E_R(\rho_I(\alpha))=
\begin{cases}
  0, & \text{if}\ \alpha <\frac{1}{d+1} \\
  S(\rho_I(\alpha)\vert\vert\rho_I(\frac{1}{d+1})), & \text{otherwise}
\end{cases}. \]

\begin{figure}[h]
	\centering
	\includegraphics[width=\linewidth]{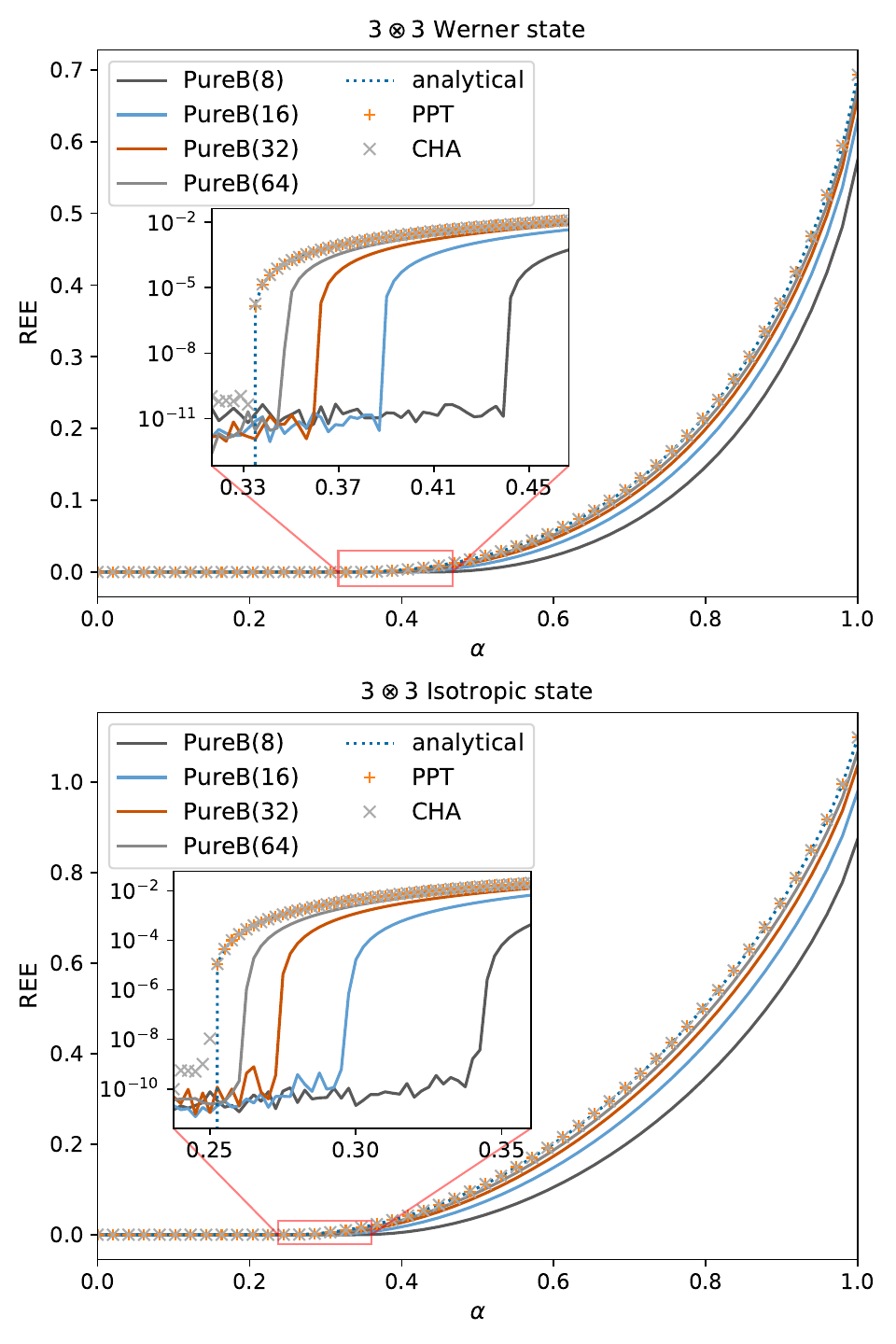}
	\caption{\label{werner_and_isotropic}Results of the REE calculated by analytical method, PPT criterion, CHA, and PureB-ext.}
\end{figure}

Firstly, we calculate the REE for $3\otimes3$ Werner states and isotropic states respectively, with the following approaches: the analytical method, PPT criterion, CHA method, and our method, PureB-ext. The results are shown in Fig. \ref{werner_and_isotropic}, which demonstrates the numerical behavior of the PureB-ext compared to others.

The results obtained by the PPT criterion, and CHA method are consistent with the analytical curve while PureB-ext provides a series of satisfactory lower bounds that approach the analytical results as the value of $k$ increases. And we can see the clear boundaries for each pure bosonic extension.

Then we compare the symmetric/bosonic extensions in QETLAB with PureB-ext. QETLAB is a MATLAB toolbox for quantum information that provides functions for determining whether a state is $k$-symmetric/bosonic extendible based on convex optimization. Here, we list the boundary values $\alpha^{\star}$ of $2\otimes 2$ Werner states across different $k$-symmetric/bosonic extendible sets, computed by different methods. We set a threshold of $\epsilon=10^{-7}$ to determine the boundaries of pure $k$-bosonic extensions by computing REE. The results are shown in Table \ref{qet}.

\begin{table}[H]
	\centering
	\caption{\label{qet}Boundary values $\alpha^{\star}$ of $2\otimes 2$ Werner states across different $k$-symmetric/bosonic extendible sets. "NA" means it's not available within the acceptable time.}
	\begin{tabular}{c|c c c c}
		\hline
		\textbf{k}  &\textbf{Analytical} &\textbf{PureB-ext} & \textbf{Bos-ext}  & \textbf{Sym-ext}  \\
		  	\hline
		5	& 0.63636 & 0.63661	& 0.63650	& 0.63650 \\
		6	& 0.61538 & 0.61206	& 0.61556	& 0.61554 \\
		7	& 0.6 & 0.60019	& 0.60016	& 0.60017 \\
		8	& 0.58823 & 0.58846	& 0.58840	& NA \\
		9	& 0.57894 & 0.57918	& 0.57919	& NA \\
		10	& 0.57142 & 0.57165	& 0.57168	& NA \\
		11	& 0.56521 & 0.56547	& 0.56548	& NA \\
		12	& 0.56 & 0.56024	& NA			& NA \\
		16	& 0.54545 & 0.54575	& NA			& NA \\
		512	& 0.50146 & 0.50175	& NA			& NA \\
		8192	& 0.50009 & 0.50040	& NA			& NA \\
		65536	& 0.50001 & 0.50034	& NA			& NA \\
		\hline
	\end{tabular}
\end{table}

According to the previous study \cite{PhysRevA.99.012332}, symmetric and bosonic extendible sets are identical when $d_B=2$. As a result, the outcomes derived using the Sym-ext and Bos-ext functions in QETLAB are consistent. The analytical values for different extensions \cite{johnson2013compatible} are also included in the table for reference. As we can see, PureB-ext effectively characterizes the $k$-bosonic extendible set, providing highly accurate boundary values (even though the threshold $\epsilon$ we choose will affect). On the other hand, QETLAB is unable to manage large dimensions, whereas our algorithm can handle tens of thousands of extensions and asymptotically approach the boundary of the separable set ($\alpha=\frac{1}{2}$).

\subsection{Bound entanglement}\label{Bound_ENT}

In this section, our algorithm is utilized to detect bound entangled states (BES), specifically PPT BES, a category for which the PPT criterion fails to yield any entanglement information. Prior research has presented a special family of two-qutrit PPT BES \cite{divincenzo2003unextendible}, as the following:
\[ \rho =\frac{1}{d_A d_B-d_{\mathcal{S}}}\mathcal{P}_{\mathcal{S}}^{\bot} , \]
where $d_A$, $d_B$ are the local dimensions of $\mathcal{H}_A \otimes \mathcal{H}_B$, $\mathcal{S}$ is a subspace spanned by some unextendible product basis (UPB) with dimension $d_{\mathcal{S}}$, and $\mathcal{P}_{\mathcal{S}}^{\bot}$ is the projector onto the orthogonal complementary space of $\mathcal{S}$.

Consider the case of $3\otimes 3$ density matrices, where we select two special BESs, $\rho_{\text{tiles}}$ and $\rho_{\text{pyramid}}$ with the respective subspaces
\[
	\begin{aligned}
		\mathcal{S}_{\text {tiles }}= & \operatorname{span}\{|0\rangle \otimes(|0\rangle-|1\rangle),|2\rangle \otimes(|1\rangle-|2\rangle), \\
		& (|0\rangle-|1\rangle) \otimes|2\rangle,(|1\rangle-|2\rangle) \otimes|0\rangle,\\ &(|0\rangle+|1\rangle+|2\rangle) 
		 \otimes(|0\rangle+|1\rangle+|2\rangle)\}\\
		\mathcal{S}_{\text {pyramid }}= & \operatorname{span}\{|\psi_i\rangle \otimes |\psi_{2i \operatorname{mod} 5}\rangle, i=0, \dots, 4 \},
		\end{aligned}
\]
where $|\psi_i\rangle = \cos(\frac{2\pi i}{5})|0\rangle + \sin(\frac{2\pi i}{5})|1\rangle+ \frac{1}{2}\sqrt{1+\sqrt{5}}|2\rangle$.

\begin{figure}[H]
	\centering
	\includegraphics[width=\linewidth]{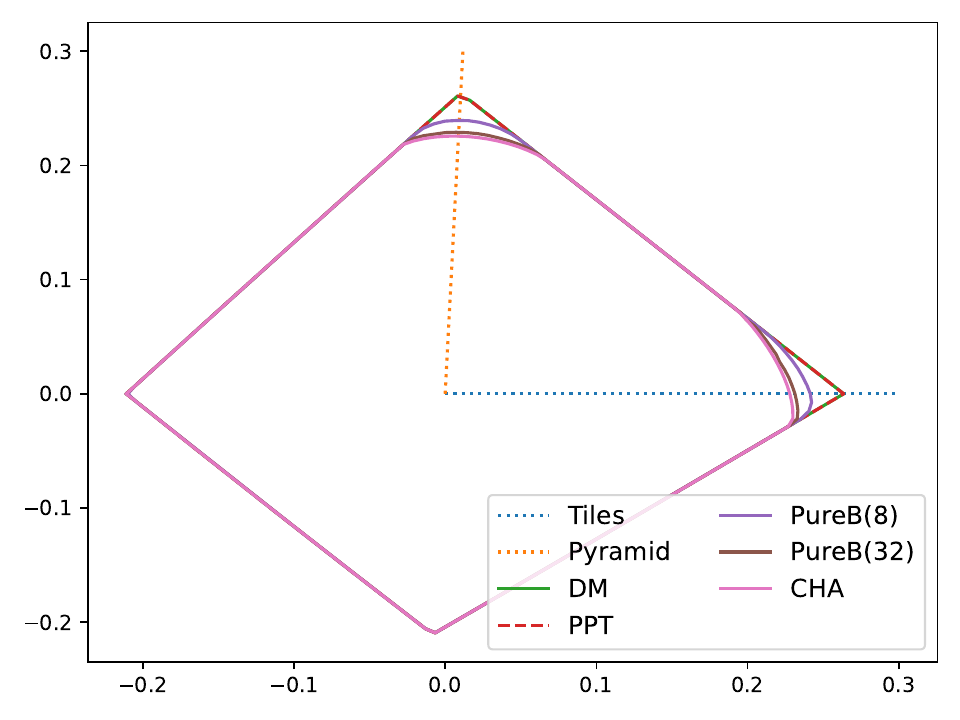}
	\caption{Two-dimensional cross section spanned by two UPB BES $\rho_{\text{tiles}}$ and $\rho_{\text{pyramid}}$ in the density matrix space. Using different methods, we can obtain different boundaries to approximate the real boundary of separable states. CHA approximates the boundary from inside, while PPT and PureB-ext approximate the boundary from outside. On this plane, the boundaries of quantum state (DM) and PPT are identical. The bound entanglement is expected to hide in the gap between CHA and PPT.}
	\label{BE}
\end{figure}

Utilizing Eq.~(\ref{eq:dm-vector-representation}), we can convert them to their vectorized forms, $\vec{\rho}_{\text{tiles}}$ and $\vec{\rho}_{\text{pyramid}}$. On the two-dimensional cross section spanned by these vectors, we delineate the boundaries obtained by different approaches, including the PPT criterion, CHA, and PureB-ext. These boundaries are the approximations of the real boundary of separable states, from inside (CHA) or outside (PPT, PureB-ext). We achieve this by identifying the boundary states $\sigma$ along different directions on this plane and then compute $||\vec{\sigma}||_2$. See Appendix \ref{app:boundary} for detailed procedures. The results are illustrated in Fig. \ref{BE}.

From the figure, it is evident that the boundaries of PureB-ext with $k=8,32$ lie between the boundaries of PPT and CHA, certifying that PureB-ext can successfully detect some bound entanglement within that region.

\subsection{Random states}

\begin{figure}[H]
	\centering
	\includegraphics[width=\linewidth]{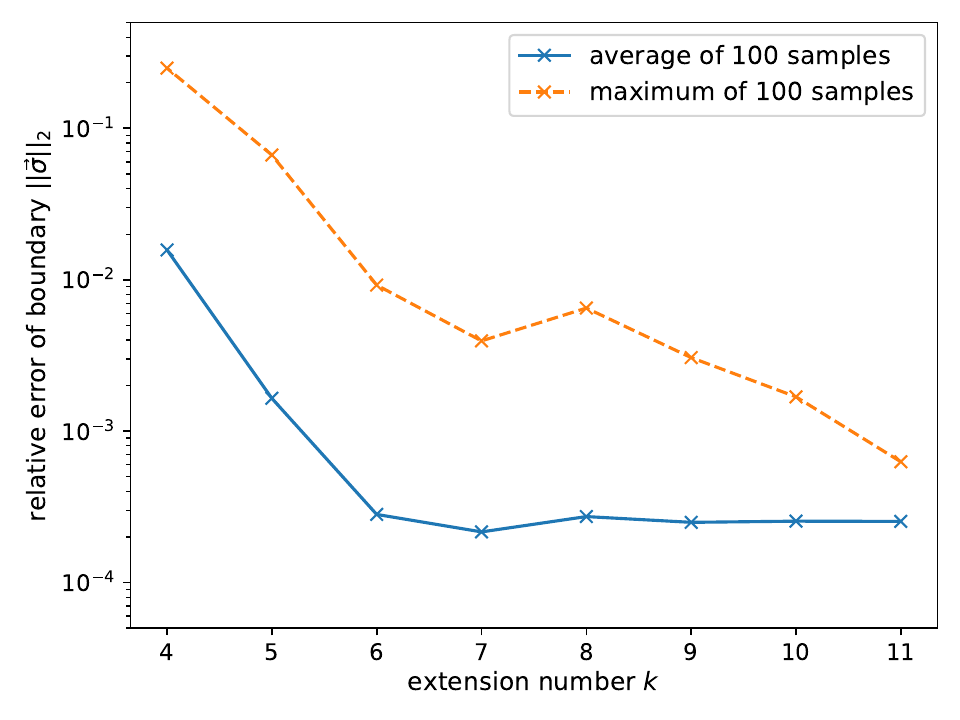}
	\caption{Relative errors of PureB-ext for computing the length of a boundary state $\sigma$, denoted by $||\vec{\sigma}||_2$. 100 random directions in $2\otimes 2$ density matrix space are selected for each extension. However, since the QETLAB cannot handle too large extension number $k$, we only plot the results for $k\leq 11$.}
	\label{fig:kext-relative-accurcy-2x2}
\end{figure}

To compute the accuracy of PureB-ext for random states, we generate 100 random directions in the $2 \otimes 2$ density matrix space, then compute the length of the boundary state in each direction for different extension numbers $k$. Subsequently, we plot the max/average relative error (assuming that QETLAB provides the exact value) in Fig.~\ref{fig:kext-relative-accurcy-2x2}. The average relative error initially decreases and then remains constant. The plateau arises because we set the converge tolerance in QETLAB around $10^{-4}$ to avoid numerical instability. These results show that our method is reliable for random states.

In Table \ref{time-compared-with-qetlab}, the related computation time of determining those boundaries is listed. The benchmark is performed on a standard laptop with AMD R7-5800H, 16 CPU cores (hyperthread enabled), and 16GB memory.  It is clear that the time required by QETLAB increases exponentially with $k$, while that required by PureB-ext increases linearly. This behavior can be attributed to the fact that QETLAB stores the full density matrix of size $2^{k+1}\times 2^{k+1}$, while PureB-ext only requires a vector of size $2(k+1)$.

\begin{table}[H]
	\centering
	\caption{Computation time (in second) of determining the boundaries of $k$-bosonic extensions in a $2\otimes2$ system. "-" means we skip these numbers of extensions in comparison since PureB-ext is nonsense if $k$ is smaller than the transition value $k^{\star}=4$ in this case. And "NA" implies it's not available within the acceptable time.}
	\label{time-compared-with-qetlab}
	\begin{tabular}{c|c c}
		\hline
		\textbf{k}   &\textbf{PureB-ext} & \textbf{QETLAB} \\
		  	\hline
        1 & - & $0.002\pm 0.019$\\
        2 & -  &$0.003\pm 0.022$\\
        3 & -  &$0.339\pm 0.564$\\
        4 & $6.312 \pm 4.276$  &$0.561\pm 0.727$\\
        5 & $7.762\pm 6.047$ &$0.707\pm 0.796$\\
        6 & $8.974\pm 6.543$ &$0.812\pm 0.835$\\
        7 & $10.358\pm 7.425$ &$1.137\pm 0.945$\\
        8 & $10.041\pm 7.121$ &$1.854\pm 1.460$\\
        9 & $9.954\pm 6.783$ &$8.816\pm 6.756$\\
        10 & $9.844\pm 6.396$ &$144.819\pm 106.711$\\
        11 & $9.911\pm 6.500$ &$2496.404\pm 1839.132$\\
        16 & $8.294\pm 5.755$ & NA\\
        64 & $7.570\pm 5.047$  & NA\\
        256 & $7.788\pm 5.104$ & NA\\
        1024 & $9.744\pm 5.715$ & NA\\
        4096 & $18.493\pm 10.055$ & NA\\
        16384 & $56.196\pm 29.903$ & NA\\
		\hline
	\end{tabular}
\end{table}
\section{Discussions}\label{Sec:discussions}

In the present study, we unveil a novel methodology named "pure bosonic extension" (PureB-ext). This innovative approach facilitates efficient characterization of the $k$-bosonic extendible set and enables precise computation of lower bounds for the relative entropy of entanglement (REE). Our numerical results suggest that PureB-ext is effective, resource-efficient, and able to handle large dimensions. In comparison to traditional convex optimization methods, PureB-ext requires relatively few parameters to optimize, which makes it outperform the bosonic extension function in QETLAB, a widely used tool for quantum entanglement. However, PureB-ext is not "silver bullet" since it is \textit{generically} exact and we may identify some special directions in the density matrix space that it struggles to handle (see Appendix \ref{tough_case}). Also, the mechanism of PureB-ext is not fully understood. We leave the investigation of these issues for future work.

The symmetric/bosonic extension proves to be a versatile tool, not merely employed to characterize the set of separable states, but also relevant to a variety of other domains. These include the quantum marginal problem \cite{Klyachko_2006}, and quantum key distribution \cite{myhr2009symmetric}, among others.

In recent work, a hybrid quantum-classical algorithm has been proposed for detecting and quantifying entanglement \cite{wang2022detecting}. Our pure bosonic extension method can also be adapted for use in such a hybrid algorithm. With the assistance of NISQ-era devices, this hybrid approach can be used to effectively handle larger dimensions. We have successfully demonstrated it for the $2 \otimes 2$ case, with the quantum circuit ansatz and numerical results provided in the Appendix \ref{Sec:hybrid}.

The set of pure states is not convex, but our research has shown that it is connected to the convex set of mixed states. There have been recent studies exploring this connection \cite{patel2021variational,bharti2022noisy}. We hope that our work will inspire further research into the non-convexity in quantum information.

\acknowledgments

We gratefully acknowledge the assistance of ChatGPT in facilitating the writing process. X-R. Zhu, C. Zhang, C-F. Cao and B. Zeng are supported by GRF grant No. 16300220.
Y-N.Li is supported by National Natural Science Foundation of China under Grant No. 12005295. Y. T. Poon is supported by the Scott Hanna Faculty Fellowship.
\appendix

\section{Calculation of $||\vec{\sigma}||_2$}\label{app:boundary}

Here, we explain how to compute the length of the boundary state $\sigma$ along the given direction $\hat{\sigma}$ for different methods we used in the main text. For simplicity, $||\vec{\sigma}||_2$ is denoted as $\beta$

\paragraph*{Boundary of the density matrix $\beta_\mathrm{DM}$} the smallest eigenvalue of the density matrix is zero,
\[\lambda_{\mathrm{min}}(\rho_0+\beta_{\mathrm{DM}}\hat{\sigma}\cdot\vec{\lambda})=0,\]
where $\lambda_{\mathrm{min}}$ denotes the smallest eigenvalue and $\vec{\lambda}$ are the Gell-Mann matrices.

\paragraph*{Boundary of the PPT set $\beta_\mathrm{PPT}$} the smallest eigenvalue of the partial transposed density matrix is zero
\[\lambda_{\mathrm{min}}(\rho_0+\beta_\mathrm{PPT}\hat{\sigma}\cdot\vec{\lambda}^{\Gamma})=0\]
where $\vec{\lambda}^\Gamma$ is short for the partial transpose of the Gell-Mann matrices with respect to one partite.

\paragraph*{Boundary of the CHA $\beta_{\mathrm{CHA}}$} the set is characterized by the convex hull approximation (CHA) of a series of pure separable state $|\psi_{A/B}^{(i)}\rangle$. To find the boundary, the following linear programming is required to solve
\begin{align*}
	\underset{\lambda_i}{\mathrm{maximize}} &\quad \beta\\
	s.t.\quad  & \rho_0+\beta\hat{\sigma}\cdot\vec{\lambda} =\sum_{i=1}^N{\lambda _i|\psi _{A}^{\left( i \right)}\rangle \langle \psi _{A}^{\left( i \right)}|\otimes |\psi _{B}^{\left( i \right)}\rangle \langle \psi _{B}^{\left( i \right)}|}, \\
	& \lambda_i\geq 0,\quad \sum_{i=1}^N \lambda_i=1
\end{align*}
with the pure states $|\psi_{A/B}^{(i)}\rangle$ updated according to an iterative strategy introduced in the previous research \cite{lu2018separability,hou2020upper}. The number of pure states is chosen to be $N=2d_A^2d_B^2$ in our calculation.

\paragraph*{Boundary of PureB-ext $\beta_{\mathrm{PureB}}$} When the density matrix $\rho$ moves from the maximally mixed state $\rho_0$ to the boundary of the density matrix set, we notice that there is a sharp and clear boundary where the relative entropy of entanglement $E_R$ given by the PureB-ext goes from zero to some nonzero value as shown in the Fig.~\ref{werner_and_isotropic}. With such an observation, we choose the density matrix with $E_R=10^{-7}$ as the signal of the PureB-ext's boundary. To find the boundary $\beta_{\mathrm{PureB}}$, we applied the binary search algorithm with the initial upper bound $\beta_{\mathrm{DM}}$ and the initial lower bound $\beta=0$.

\section{Challenging scenarios for PureB-ext}\label{tough_case}

\begin{figure}[h]
	\centering
	\includegraphics[width=\linewidth]{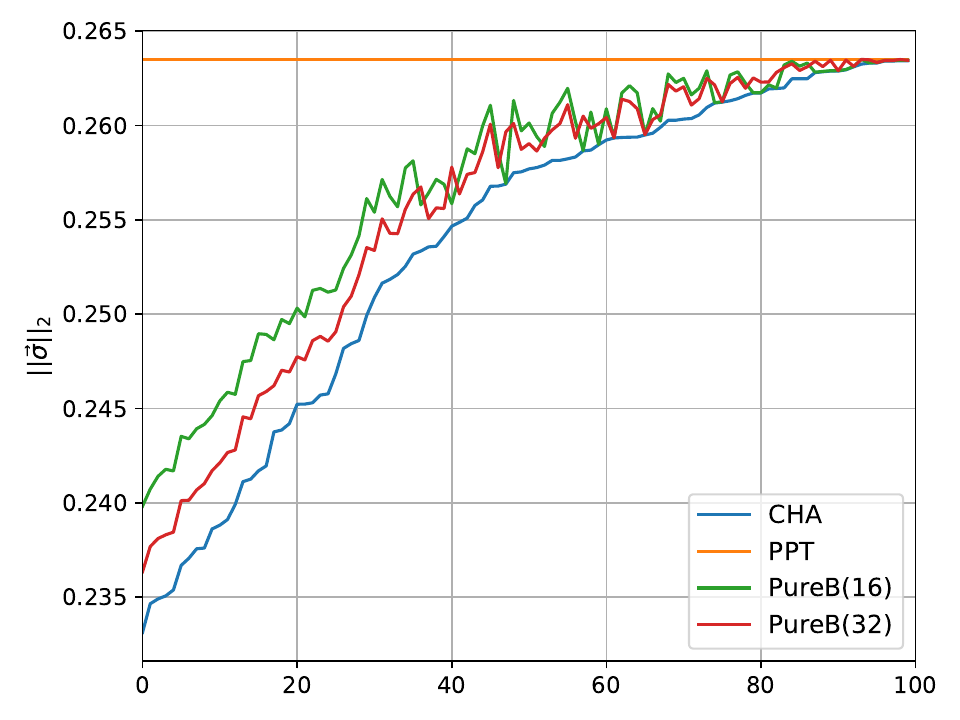}
	\caption{$||\vec{\sigma}||_2$ given by different PureB-ext, CHA, and PPT for randomly generated 100 samples from the six-parameter family. The horizontal axis is the index of the samples sorted by the values given by CHA.}
	\label{sixparam_random}
\end{figure}

Here, we consider the higher dimension where there exists bound entanglement. Bound entanglement is a suitable testing ground to verify our method's validity. 100 samples are randomly generated from the six-parameter family in $3\otimes3$ \cite{divincenzo2003unextendible}, which gives special directions in the density matrix space. The lengths of the associated boundary states $||\vec{\sigma}||_2$ for different approaches are calculated, including CHA, PPT, and PureB-ext in Fig.~\ref{sixparam_random}. We rearrange the samples according to the values calculated by CHA. The values obtained by PPT are identical due to the special property of the six-parameter family. So you can observe a straight line in the figure below. Meanwhile, our method PureB-ext gives satisfactory values among the most samples. As expected, we can get more accurate results to approximate the separable set by increasing the number of extensions. However, there are some overlaps among the different PureB-ext. These directions might be the degenerate cases, where we cannot find a pure pre-image for the boundary state of $\bar{\Theta}_k$ (or due to some convergence problems from numerical instability which we cannot confirm). So there might be a quasi-inclusion relation between PureB($k$) and PureB($k+1$) instead of a strict inclusion relation. In practice, we only observed the "bad" directions in some bound entangled corners where the gap between PPT and CHA is small. For generic directions, PureB-ext is still effective and efficient.

\section{Hybrid quantum-classical algorithm}\label{Sec:hybrid}

Although PureB-ext only requires pure states with a much smaller dimension than density matrices, the calculation for larger dimensions could be time-consuming and inefficient for classical computers. Here, we consider a symmetric variational quantum circuit to produce a pure bosonic state, as shown below.
\begin{figure}[H]
	\centering
	\includegraphics[width=\linewidth]{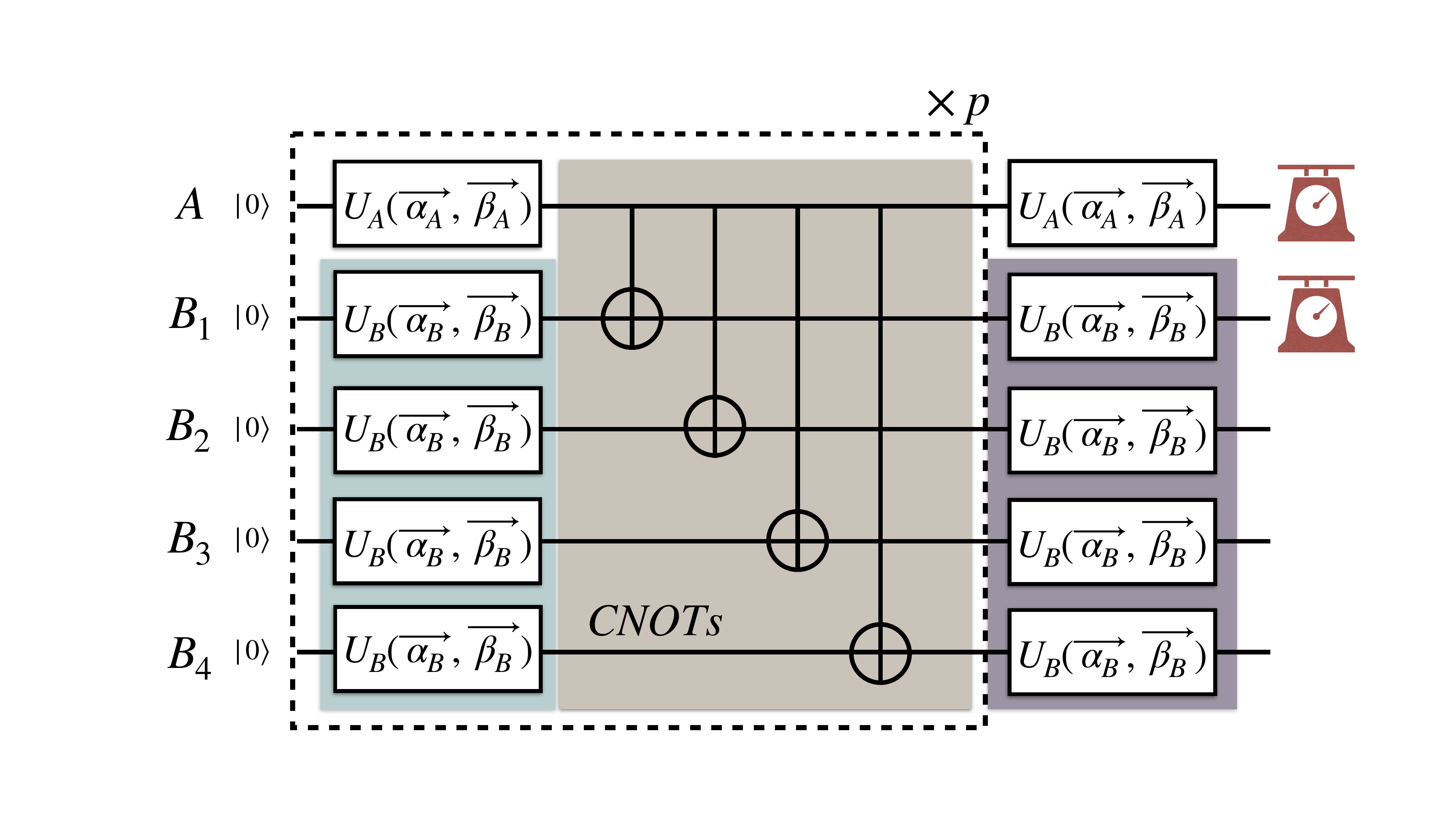}
	\caption{\label{qc}The quantum circuit for a pure 4-bosonic extension.}
\end{figure}
We introduce two kinds of quantum gates for qudit:

\paragraph*{Single-qudit gate} Weyl group generators $X$ and $Z$ are used to parameterize the unitary matrix for qudit
\[ U(\vec{\alpha},\vec{\beta})=\prod_{i}X^{\alpha_i}Z^{\beta_i},\]
and any unitary matrices can be approximated by a long enough sequence $\vec{\alpha},\vec{\beta}$ \cite{PhysRevA.68.062303}.

\paragraph*{Double-qudit CNOT gate} a generalized version of CNOT for qudit is defined as
\[ \tilde{CX}\vert x\rangle\vert y\rangle = \vert x \rangle \vert (y-x)\% d_B \rangle.\]
To add the interaction between system A and symmetric system B, a series of CNOT gates are applied on every pair of control qudit $A$ and target qudit $B_i$ as below,
\begin{align*}
	\bigotimes_{i=1}^{k}& \tilde{CX}^{\left(A B_{i}\right)}|x\rangle\vert D_{w_{0}, w_{1}, \cdots, w_{d-1}}^{k}\rangle \\=&\left\{\begin{array}{l}
		|x\rangle\vert D_{w_{0}, w_{1}, \cdots, w_{d-1}}^{k}\rangle \quad x=0 \\
		|x\rangle \vert  D_{w_{1}, \cdots, w_{d-1}, w_{0}}^{k}\rangle \quad x=1 \\
		|x\rangle\vert D_{w_{2}, \cdots, w_{d-1}, w_{0}, w_{1}}^{k}\rangle \quad x=2 \\
		\vdots
		\end{array}\right. .
\end{align*}
Such a symmetric CNOT gate changes the symmetric basis of $B_1...B_k$ according to the state of control qudit in $\mathcal{H}_A$. Hopefully, any pure bosonic state $\vert \psi_{AB_1...B_k}\rangle$ can be generated with enough layers using these two kinds of single-qubit and double-qudit gates.

We will show the relative entropy of entanglement solved by the PureB-ext on a variational quantum qubit circuit. For qubit case, the Weyl group generators $X,Z$ become the canonical Pauli matrices and the effective length for the $\vec{\alpha},\vec{\beta}$ in the single-qubit gate is $1$ for that the Pauli matrix $\sigma_z^{(0)}$ or $\sigma_x^{(1)}$ is commutative with the CNOT gate. The effective number of parameters in each layer is $4$ for the circuit in Fig. \ref{qc}, $2$ parameters for $U_A$, and $2$ for $U_B$. Compared with the number $4(k+1)-1$ of parameters in the classical version of the PureB-ext, where $4$ counts both real and imaginary parts and $-1$ is for the normalization constraint, a direct conjecture is that the minimum number of layers for the quantum version of PureB-ext is $(k+1)$.

\begin{figure}[H]
    \centering
    \includegraphics[width=\linewidth]{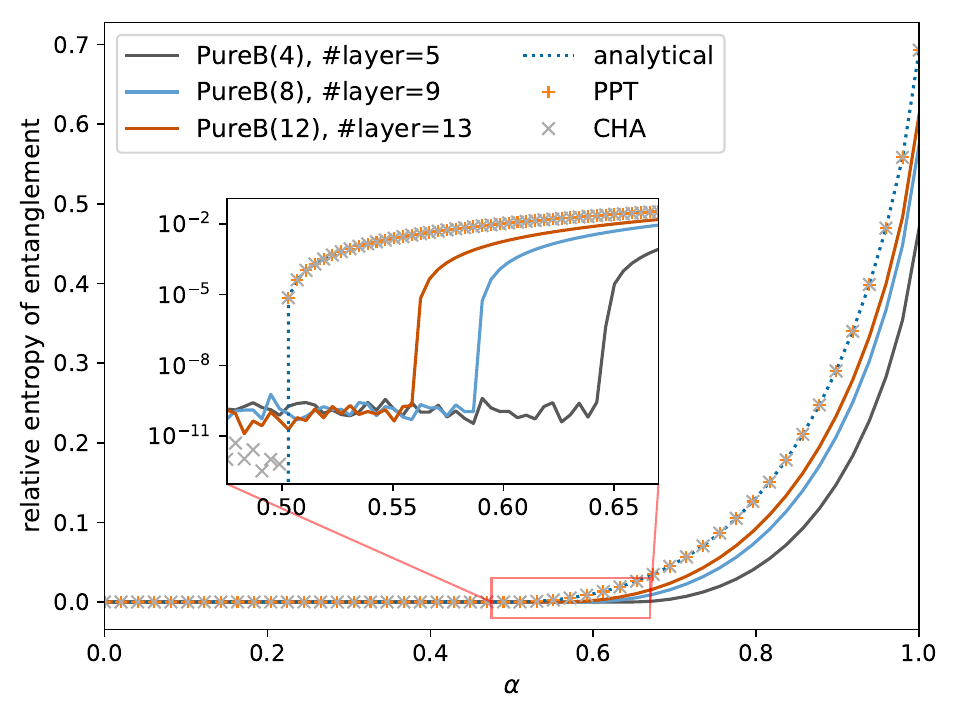}
    \caption{\label{nisq}The numerical simulation of PureB-ext on a symmetric variational quantum circuit. Each PureB($k$) has $k+1$ layers for optimization. The clear and sharp transitions near the boundaries also emerge in the quantum version.}
\end{figure}

In Fig. \ref{nisq}, the relative entropy of entanglement for the Werner-2 is solved by the PureB-ext with $k=4,8,12$ parameterized on quantum circuits. We could see that the variational quantum circuits also give a sharp and clear boundary where the entropy changes from zero to some nonzero value. As $k$ increases, the boundary of the PureB-ext approaches the analytical boundary $\alpha=0.5$.

\section{Data Availability}

Both the code and data for our project have been made publicly available. Our open-source repository can be accessed at \cite{github-repo}.

\bibliography{manuscript.bib}
\end{document}